\documentclass[
preprintnumbers,
 amsmath,amssymb,
 aps,
nofootinbib,
floatfix]{revtex4-2}

\usepackage{xcolor}

\usepackage{url}
\usepackage[hyperindex,breaklinks]{hyperref}
\hypersetup{linkcolor=blue, citecolor=red, colorlinks=true}

\usepackage{braket}
\usepackage{physics}
\usepackage{mathtools}
\usepackage{amsfonts}

\usepackage{graphicx}
\usepackage{dcolumn}
\usepackage{bm}
\usepackage{diagbox}

\usepackage{float}

\usepackage{placeins} 

\usepackage{soul}

\begin{document}

\title{Quantum walks  with  spatiotemporal fractal disorder}

\author{$^{\star}$Marcelo A. Pires$^{1}$}
\thanks{marcelo.pires@delmiro.ufal.br}

\author{$^{\star}$Caio B. Naves$^{2,3}$}
\thanks{caio.naves@alumni.usp.br}

\author{Diogo O. Soares-Pinto$^{2}$}
\thanks{dosp@ifsc.usp.br}

\author{S\'{\i}lvio M. \surname{Duarte~Queir\'{o}s}$^{4,5}$}
\thanks{sdqueiro@cbpf.br} 

\affiliation{
$^{1}$Universidade Federal de Alagoas, 57480-000, Delmiro Gouveia - AL, Brazil
\\
$^{2}$Instituto de F\'{\i}sica de S\~ao Carlos, Universidade de S\~ao Paulo, CP 369, 13560-970, S\~ao Carlos - SP, Brazil
\\
$^{3}$Department of Physics, Stockholm University, AlbaNova University Center, 106 91 Stockholm, Sweden
\\
$^{4}$Centro Brasileiro de Pesquisas F\'{\i}sicas, Rua Dr Xavier Sigaud 150, 22290-180, Rio de Janeiro - RJ, Brazil
\\
$^{5}$National Institute of Science and Technology for Complex Systems, Brazil
}

\date{\today}

\begin{abstract}
We investigate the transport and entanglement properties exhibited by quantum walks with coin operators concatenated in a space-time fractal structure. Inspired by recent developments in photonics, we choose the paradigmatic Sierpinski gasket. The 0-1 pattern of the fractal is mapped into an alternation of the generalized Hadamard-Fourier operators. In fulfilling the blank space on the analysis of the impact of disorder in quantum walk properties -- specifically,  fractal deterministic disorder --, our results show a robust effect of entanglement enhancement as well as an interesting novel road to superdiffusive spreading with a tunable scaling exponent attaining effective ballistic diffusion.
Namely, with this fractal approach it is possible to obtain an increase in quantum entanglement without jeopardizing spreading.  Alongside those features, we analyze further properties such as the degree of interference and visibility. The present model corresponds to a new application of fractals in an experimentally feasible setting, namely the building block for the construction of photonic patterned structures.
\end{abstract} 

\maketitle

\footnotetext{$^\star$These authors contributed equally to this work.}


\section{\label{sec:intro}Introduction}

Scale-invariance, i.e. the intuitive sense of indistinguishability between the overall shape of several natural systems and a portion of them, has marveled humankind since a long time. Ultimately, this paved the way to the concept of fractals -- either self-similar or self-affine --, i.e. objects whose capacity and Lebesgue covering dimension are different from one another. Besides its inherent beauty, fractality has found its place at the first tier of science and technology~\cite{mandelbrot1983fractal,vehel2012fractals,bunde2013fractals}. The endeavor to set forth fractal theory applications in the development of new scale-invariant structures can be gauged by the amount of literature that has been produced in recent years. For instance, it was shown it is possible to use molecular self-assembly to engineer an artificial nanometer-scale Sierpinski hexagonal gasket~\cite{newkome2006nanoassembly}, structures such as Peano, Greek cross and the Vicsek constructions to build stretchable electronic devices with unusual mechanics~\cite{fan2014fractal}. On the other hand, fractals went deep into the quantum realm by the assembling of artificial architectures of electrons with fractal geometry~\cite{shang2015assembling,kempkes2019design}.

Such studies allowed the experimental investigation of subjects that have been studied only from a theoretical perspective for long. In this work, we present a theoretical investigation of position- and time-dependent quantum walks with fractals that are feasible to experiments with newly developed photonic architectures.

Quantum Walks (QW for short)~\cite{aharonov1993quantum} are the celebrated proxy lattice models for studying wavepacket spreading. Within the scope of disorder-free QWs, it is well-known the emergence of interesting properties such as ballistic spreading with a non-gaussian bimodal probability distribution that are both very distinct from the classical random walk.
Still, a richer phenomenology comes up when new ingredients are introduced
in the coin~\cite{vieira2013dynamically,chandrashekar2012disorder,chen2016defect,ampadu2013return,pires2020parrondo,pires2021negative,fang2023maximal} or step~\cite{lavivcka2011quantum,di2018elephant,
pires2019multiple,pires2020quantum,ahmad2020randomizing,naves2023quantum,naves2022enhancing} operators.

On the one hand, temporal disorder in QWs can be implemented by breaking the time independence of the walk operators whilst keeping their spatial translation invariance~\cite{pires2020parrondo,banuls2006quantum,dhar2019diverse,xue2015experimental,romanelli2009driving,machida2010limit,panahiyan2019simulation}. 
On the other hand, spatial disorder in QWs can break the lattice constancy of the walk operators without changing their time invariance~\cite{konno2010localization,Li2013,Zhang2014,cantero2012one,buarque2019aperiodic,ahmad2020one}.
A third type of disorder -- the focus of our work -- involves breaking both space and time constancy of the walk operators. 
In a comprehensive study~\cite{ahlbrecht2012asymptotic}, it was shown that spatio-temporal randomness embedded in the coin operator promotes the transition to a classical-like spreading in quantum walks. Other works -- namely Refs.~\cite{chandrashekar2012disorder,vieira2014entangling} -- confirmed such findings, where it was noticed that random spatio-temporal disorder leads to a slowing down in the wavepacket spreading. 
In Ref.~\cite{montero2017quantum}, it was found that carefully devised nonrandom space-time disorder can be used as a probability distributions universal generator. Particularly, in Ref.~\cite{montero2016classical}, it was introduced a method to create spatio-temporal dependence in the coin operator that is able to display an exact classical-like binomial distribution without randomness. Those works naturally lead to the question of what dynamics should emerge if the coin operator is embedded in a deterministic space-time disorder tailored with nontrivial patterns such as fractals.

Although fractals can emerge as the output of a QW dynamics~\cite{shikano2010localization,cedzich2020singular}, we are going to deal with fractals the other way round, namely as the input in the modelling process and relate it to previous works as follows: in Ref.~\cite{agliari2010quantum,berry2010quantum,patel2012search,tamegai2018spatial,sato2020scaling} the focus was the problem of quantum search and in Ref.~\cite{xu2020shining}
it was conducted an experimental work on continuous-time quantum walks in fractal photonic lattices whereas in Ref.~\cite{lara2013quantum} the goal was to study the scaling of the spreading by considering a flip-flop shift operator with a four-dimensional Grover coin. Complementary, the authors of Ref.~\cite{andrade2018discrete} studied a discrete-time quantum walk (DTQW) with a two-dimensional coin operator with space dependence prescribed by the Cantor set and Sierpinski fractals were considered in a protocol of a continuous-time quantum walks (CTQWs)
~\cite{agliari2008dynamics,darazs2014transport}.
Explicitly, we take a different road by considering a theoretical proposal very close to recent experimental setups by considering a discrete-time QW with two-state coins where the fractals model a spatio-temporal alternation between generalized Hadamard and Fourier coin operators.


\section{Model}
\subsection{1D+1 quantum walk}
At a given time $t \in \mathbb{N}$ we can write the full wave function $\Psi_t$ as
\begin{align}
\Psi_t 
=
\sum_{x \in \mathbb{Z}} 
\ket{x} 
\otimes
\psi_t(x)
\label{Eq:psi_geral}
\end{align}
\begin{align}
\psi_t(x)  = \psi_t^U(x) 
\ket{U}
+
\psi_t^D(x) 
\ket{D},
\label{Eq:psi_geral2}
\end{align}
where $\psi_t^{U,D}(x) $ are the time- and site-dependent amplitudes of probability associated with the internal degree of freedom  $ c=\{U,D\}$. 

The temporal evolution proceeds with the iterative application of the operator $\widehat W$ as 
\begin{align}
\Psi_t  \xrightarrow{ \widehat W_t} \Psi_{t+1},
\label{Eq:WSC1}
\end{align}
\begin{align}
\widehat W_t = \widehat{T}(\widehat{R}_t\otimes \mathcal{I}_\mathbb{Z}),
\label{Eq:WSC2}
\end{align}
with the identity operator $\mathcal{I}_\mathbb{Z}=\sum_{x \in \mathbb{Z}} \ketbra{x}$. In addition: 
\begin{itemize}
    \item the coin operator that leaves each internal state $\{U \text{ or } D\}$ into a weighted superposition $\widehat{R}_t = \sum_x \ketbra{x} \otimes \widehat{R}_t(x)$, with
  \begin{align}
    \widehat{R}_t(x): 
    \begin{cases}
     |x, U \rangle \rightarrow 
     c_{UU}(x,t) |x, U \rangle +
     c_{DU}(x,t) |x, D \rangle
      \\
     |x, D \rangle \rightarrow 
     c_{UD}(x,t) |x, U \rangle +
     c_{DD}(x,t) |x, D \rangle
    \end{cases} ,
    \label{eq:coinpos}
  \end{align}
   where $c_{ij}(x,t)$ are the elements of the rotation matrices playing the role of quantum walk coins used in our study:  the Hadamard coin, $\widehat C_{H}$ , and the Fourier coin, $\widehat C_{F}$, mathematically described by
\begin{equation}
\widehat C_{H}
= 
\begin{pmatrix}
\cos \theta_H &   \sin \theta_H \\
 \sin \theta_H  & -\cos \theta_H
\end{pmatrix}
,
\quad\quad
\widehat C_{F}
= 
\begin{pmatrix}
\cos \theta_F &  i \sin \theta_F \\
i \sin \theta_F  & \cos \theta_F
\end{pmatrix} ;
\label{Eq:C-explicito}
\end{equation}

    \item  the spin-dependent displacement operator that splits the wavepacket towards  $\pm x$
  \begin{align}
    \widehat{T}: 
    \begin{cases}
     |x, U \rangle \rightarrow |x+1, U \rangle 
      \\
     |x, D \rangle \rightarrow |x-1, D \rangle 
    \end{cases} .
    \label{Eq:T-explicito}
  \end{align}

\end{itemize}

\subsection{The Sierpinski gasket concatenation of optical elements}
As already mentioned, herein we centre our attention on the impact of one of the canonical instances of self-similarity, the Sierpinski gasket (SG). It corresponds to a fractal with (equilateral) triangle external contour shape obtained by means of the recursive division of a first triangle into ever smaller triangles. Geometrically, it corresponds to starting with an equilateral triangle that is divided into four congruent smaller equilateral triangles with half the size and the removal of the central triangle. Analyzing from a zooming out perspective, we understand that in doubling the size of a triangle the Sierpinski generator creates three replicas of the previous triangle. That allows finding the fractal dimension of the pattern,
\begin{equation}
d_f = \frac{\ln 3}{\ln 2} = 1.58\ldots,
\end{equation}
which is less than the Lesbegue covering dimension $D = 2$ but greater than the dimension of a line, $d=1$.

\begin{figure}[h]
\centering
\includegraphics[scale=0.6]{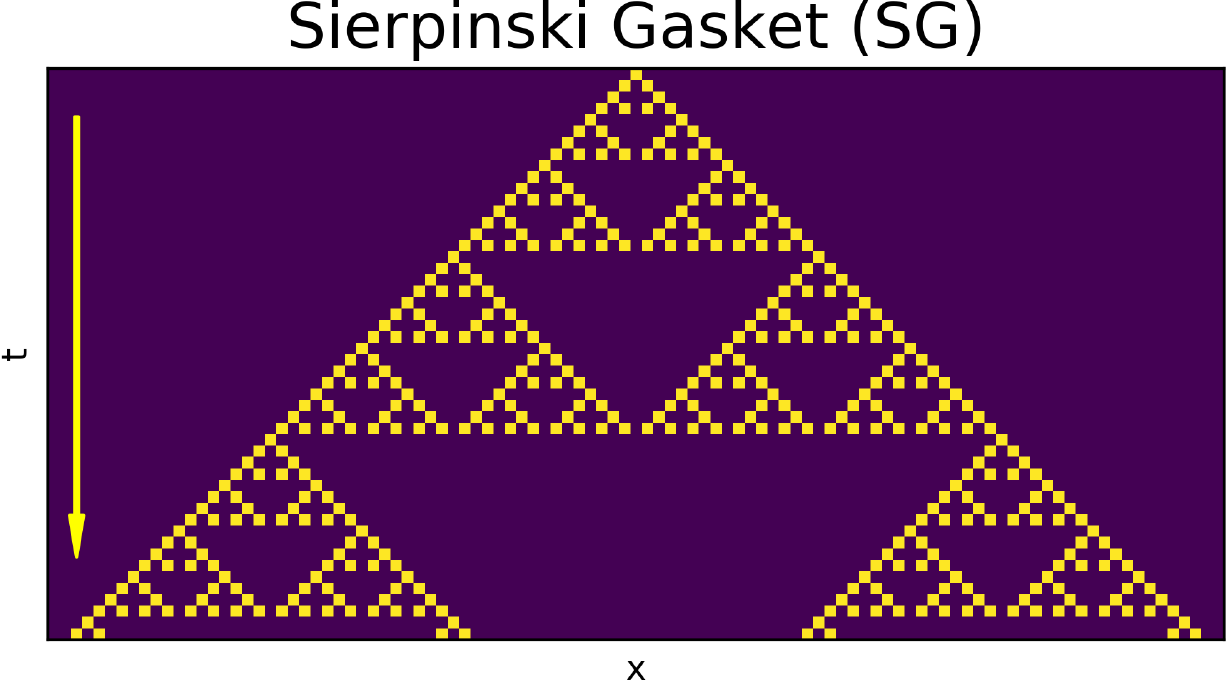}
\caption{Sierpinski gasket (SG). In this binary carpet yellow indicates the value $1$ that corresponds to the application of the operator $\widehat C_{H}$. The other sites inside the largest yellow triangle  indicate the value $0$ that corresponds to the application of the operator $\widehat C_{F}$. In this way it is possible to build a fractal assembly of optical elements.}
\label{fig:sg-sgl}
\end{figure}

Scale invariance mainly splits into two classes: when we are dealing with homogeneous dimensions and when the fractal spans through different dimensions such as space and time. In the former case we refer to self-invariance whereas in the latter we are dealing with self-affinity.
In respect thereof, the Sierpinski gasket is obtained not only using geometrical arguments, as we have mentioned, but it is the space-time outcome of several dynamical models as well. Besides the celebrated Rule~90 of cellular automata~\cite{wolfram1984rule90}, as well as a plethora of other algorithms~\cite{ettestad2019distinguishing}, the space-time SG can be obtained by means of the next fast modular arithmetic rule~\cite{nagler20051}
\begin{equation}
b_t(x) = \left[ b_{t-1}(x-1) +   b_{t-1}(x+1) \right] \mod 2
\label{eq:fractal},
\end{equation}
where $b_t(x)$ is a site- and time-dependent binary $\{0,1\}$ variable. At first glance, the Eq.~(\ref{eq:fractal}) seems to be linear, but a closer look reveals that the modular arithmetic that constrains $b_t(x)$ to either $0$ or $1$ introduces nonlinearity.
In Fig.~\ref{fig:sg-sgl}, we present $t=50$ generations of the SG. From Eq.~(\ref{eq:fractal}) it is clear that inside the cone $-t\leq x \leq t$ there is only $0-1$. 

We are now in the condition of using the self-similarity seasoning of our work into the quantum walk dynamics by defining the quantum operator
\begin{equation}
\widehat{R}_t(x) \equiv b_t(x)\widehat C_{H} + (1-b_t(x))\widehat C_{F}.
\label{eq:coinop}
\end{equation}
This equation reads as whenever a given site has $b_t(x)=0$ we set $\widehat C_{F}$ and whenever $b_t(x)=1$ we set $\widehat C_{H}$.
That establishes an important difference between our model and other photonic transport work; specifically, in the system we study the fractal element resides within the coin operator selector whereas in other cases the fractal element has to do with the physical geometry of the grid (see e.g. Ref.~\cite{xi2021network}).
Overall, we understand the present model as a space-time disordered model and such traits will be taken into consideration in the interpretation of the results.

\begin{figure*}[t]
\centering
\includegraphics[width=0.99\linewidth]{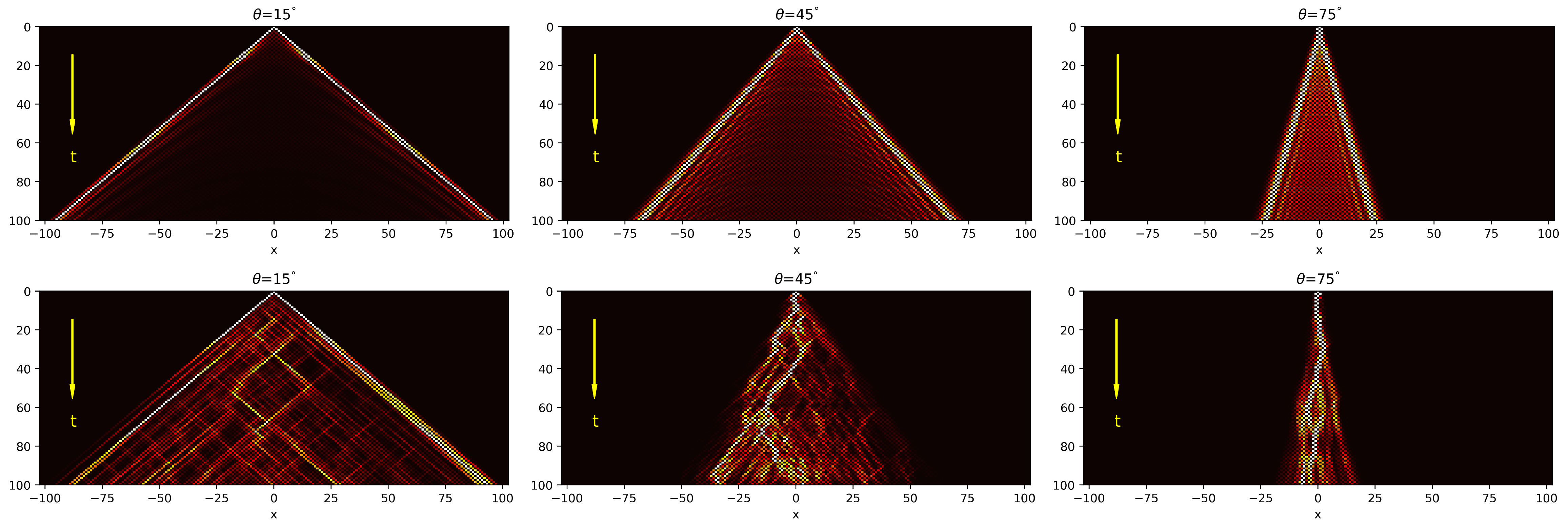}
\caption{Space-time evolution of the normalized probability distribution $P_t(x)/P_t^{max}$
where the brightness denotes the magnitude of this quantity. Quantum carpets for the disorder-free setting (upper) disordered cases (bottom). We set with $t_{max}=100$ and $\theta=\{15^o,45^o,75^o\}$. }
\label{fig:ca-qw-pxt}
\end{figure*}

With respect to the initial condition, we have placed the starting point of the QW at the same initial location of the seed in the fractals. For the boundary conditions, we have established that for each $t$ we worked with an augmented chain of positions so that the quantum walker was not able to reach the boundaries. Instances of the quantum walk as established by our dynamics are provided in Fig.~\ref{fig:ca-qw-pxt}.

Taking a friendly experimental setup, we consider the proposals in Refs.~\cite{schreiber2010coin,wang2018dynamic}, where the role of the coin operators is played by beam splitters and the internal degree of freedom $c=\{U,D\}$ is the polarization of the photons. While $\widehat C_{F}$ leaves each state in a superposition with an additional phase, $\widehat C_{H}$ produces a superposition of states without such extra phase.
If $\theta=\pi/4$, we recover the standard Hadamard and Fourier coin operators that can be implemented with an unbiased beam splitter.

\begin{figure}[h]
\centering
\includegraphics[width=0.5\textwidth]{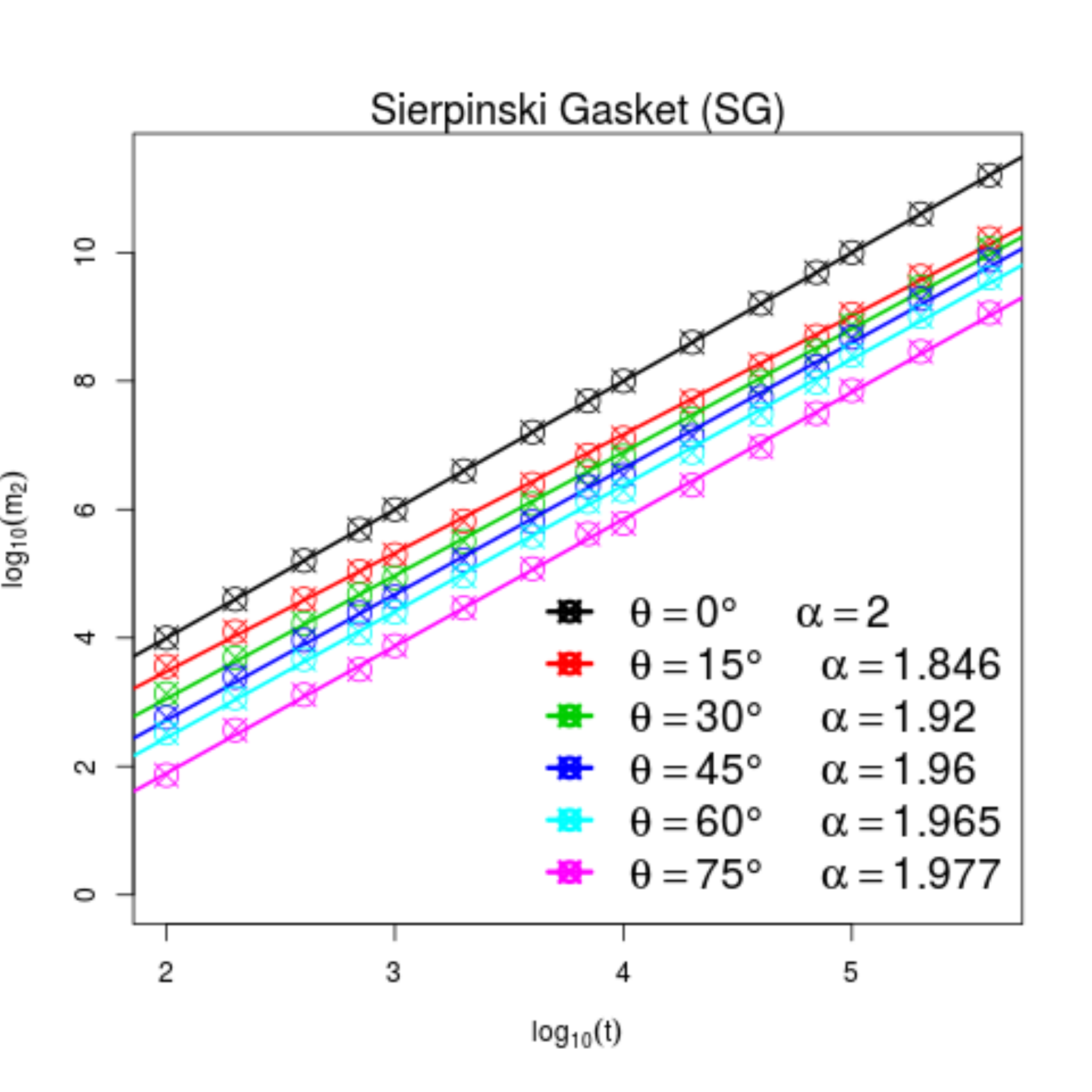}
\caption{Temporal behaviour of the spreading measure $m_{2}$ in scale $\log _{10}x\log _{10}$. Data points are the results from simulations and the lines are linear fittings.}
\label{fig:x2}
\end{figure}

\begin{figure}[h]
    \centering
    \includegraphics[width=0.5\textwidth]{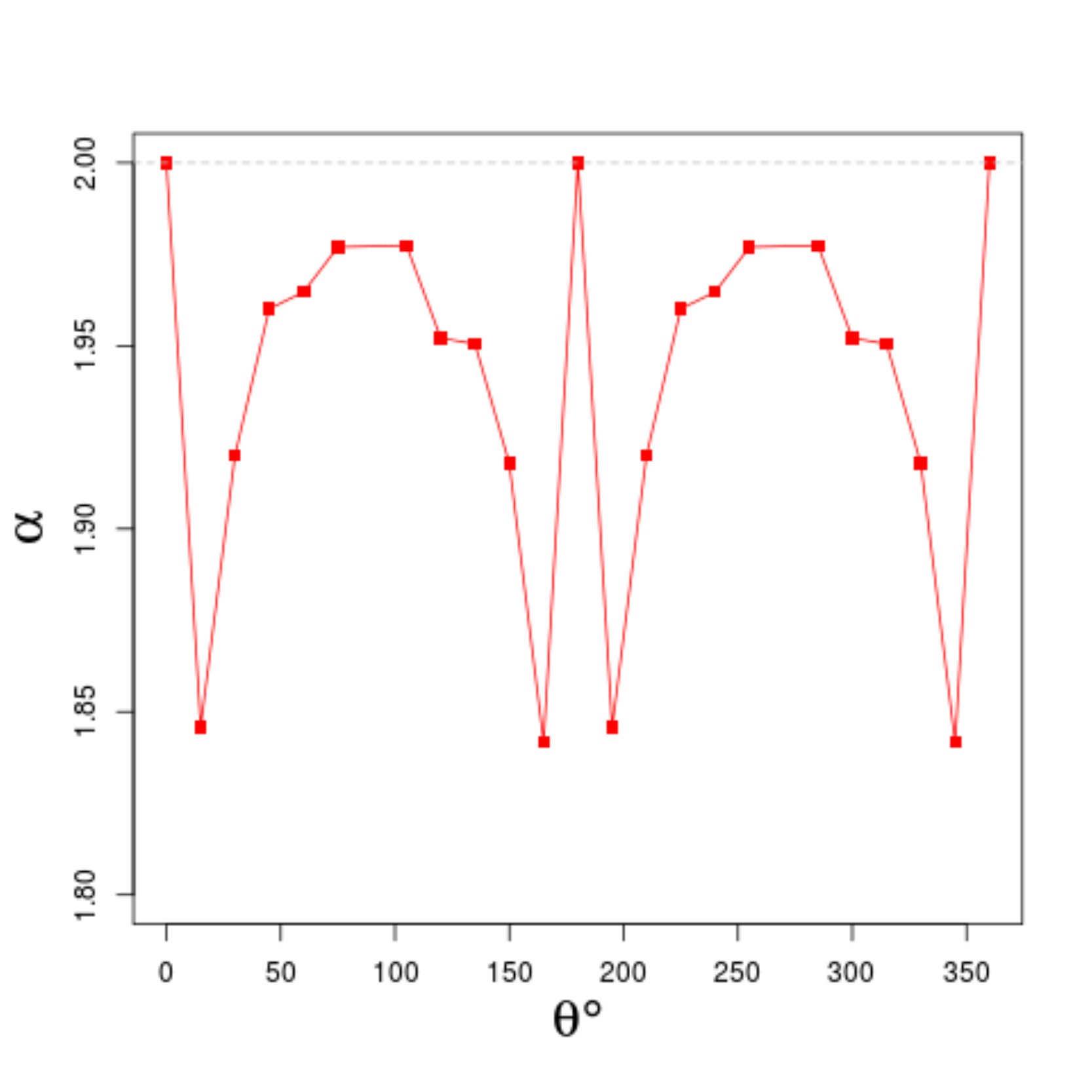}
    
    \caption{Diagram $\alpha$ vs $\theta$. We use $t_{max}=5\times 10^5$ to estimate $\alpha$ from $m_2  \sim t^\alpha$. The case $\theta=m\pi/2$ with $m=\{1,3,5\}$ is not shown because it leads to a bounded behaviour that leads to $\alpha=0$ (non-spreading) corresponding to specious localization.}
    \label{fig:alpha_theta}
\end{figure}

\section{Results}

In this section, we will present our results regarding the transport properties of the quantum walk, as well as the degree of interference, entanglement entropy, and trace distance between time-consecutive coin states.

\subsection{Spreading}

At a given instant $t$, we compute the probability
\begin{equation}
P_t(x) = |\psi_{t}^{D}(x)|^2 + |\psi_{t}^{U}(x)|^2\mbox{ .}
\end{equation}
That is of experimental interest for it allows obtaining preliminary assessment of the wavepacket transport properties. For each $t$, we also have computed $P_t^{max}$ that is the maximum of $P_t(x)$ over the chain. 

In Fig.~\ref{fig:ca-qw-pxt}, we plot the normalized profile $P_t(x)/P_t^{\max}$ for the typical angles $\theta=\{15^o,45^o,75^o\}$, where we immediately note that the standard disorder free QW (upper panel) already displays a triangle-like shape. That signals the overwhelming majority of the flux of probability is near the edges. Such a feature accommodates well our proposal for implementing QW through the Sierpinski triangle (ST).  In the disorder-free setting, we also see that the smaller the value of $\theta$, the broader the wavepacket. That is because the mathematical structure of the coin operators favors the terms $ c_{UU}$ and $ c_{DD}$ in Eq.~(\ref{eq:coinop}), which are related to spreading. However, this property is modified when the flux of probability passes through the SG. The profiles of $P_t(x)$ are neither bimodal-like distribution nor a Gaussian-like shape. 
In other words, the intrinsic nonlinearities of the fractals leaves peculiar fingerprints in the wavepacket. Even though the SG is symmetric, there are clear asymmetries in the $P_t(x)$ due to the alternation between the coin operators that give different phases to the local spinor Eq.~(\ref{Eq:psi_geral2}). 

The qualitative information obtained by the visual inspection of the quantum carpets presented Fig.~\ref{fig:ca-qw-pxt} is quantitatively boosted by computing the second statistical moment 
\begin{equation}
m_2(t) = \overline{x^2}_t =\sum_x x^2P_t(x) ,
\end{equation}
whence we compute the scaling exponent
\begin{equation}
\alpha = \lim_{t \to \infty} \frac{\log m_2(t)}{\log t},
\end{equation}
that defines the sort of diffusion the system presents. For $0<\alpha < 1$, a system is sub-diffusive; $\alpha = 1$, corresponds to standard diffusion (similar to the classical random walk); when $1<\alpha<2$, it exhibits superdiffusion; for $\alpha=2$, we have ballistic diffusion like the standard QW and for $\alpha>2$ a system runs in a superballistic regime with the particular value $\alpha = 3$ corresponding to hiperballistic diffusion~\cite{di2018elephant}.

In Fig.~\ref{fig:x2}, we understand that the nonrandom fractal disorder is not able to decrease the scaling exponent to the level of the diffusive behavior $\alpha=1$, as happens for random spatiotemporal assembly of coin operations~\cite{ahlbrecht2012asymptotic}. Therefrom, we perceive a nonmonotonic and nonsmooth dependence of $\alpha$ with $\theta$ that arises from substantial enhancement of the interference between the paths along the time evolution caused by the fractal infrastructure as previously shown in  Fig.~\ref{fig:ca-qw-pxt}. This nonmonotonicity is an intriguing feature if we take into account the overall regularity in the patterns through the SG.

The  behavior of $\alpha$ with $\theta$ is shown in Fig.~\ref{fig:alpha_theta}.  
In the standard QW it is well-known that changing $\theta$ alters the particular features of $m_2(t)$ whilst keeping the scaling exponent $\alpha$ invariant. Here, our results show that by tuning $\theta$ it is possible to slightly change (and with statistical significance) the level of superdiffusivity of the quantum walker, still below the ballistic regime though. This novel result -- the dependence of $\alpha$ with $\theta$ -- is absent in all the previous endeavors working with coined QWs on fractals~\cite{agliari2010quantum,berry2010quantum,patel2012search,tamegai2018spatial,sato2020scaling,lara2013quantum,andrade2018discrete,agliari2008dynamics,darazs2014transport}. Indeed, with our proposal it is possible to adjust $\theta = m \, \pi  \,\,\,(m \in \mathbb{N})$ in order to obtain a ballistic spreading, $\alpha=2$. On the other hand, minimal superdiffusion of the wavepacket is achieved when $\theta = m \, \pi \pm \pi/12.$

Our setup can be interpreted from the perspective of the physics of disordered systems~\cite{bunde2012fractals,bunde2013fractals}. Space-time disorder is not only of theoretical interest, but it has been shown in  optical systems that temporally fluctuating spatial disorder is able to produce a remarkably new phenomenology as well~\cite{levi2012hyper}. The results in Figs.~\ref{fig:x2} and \ref{fig:alpha_theta}  highlight the emergence of a new mechanism to access the superdiffusive regime, which is an important class of anomalous diffusion on its own~\cite{oliveira2019anomalous}.

On the whole, the results depicted in Figs.~\ref{fig:ca-qw-pxt}-\ref{fig:alpha_theta} show that a fractal assembly of coin operators leads to a rich phenomenology in terms of wavepacket spreading, even when the coin operators differ only by mere phase factors as explicit in the operators $\widehat C_{H}$ and  $\widehat C_{F}$ shown in  Eq.~(\ref{Eq:C-explicito}).

\subsection{Degree of interference}

 As a means to study with more detail the interference effects between the different paths that come up with different dynamics of the walker in this setup, we have also investigated the degree of interference. The degree of interference is defined as the norm of the quantity responsible for the interference in the visibility pattern. The visibility, in turn, suitable for probability waves is defined as 
\begin{equation}
        \mathcal{V}(x,t) \equiv \frac{\mbox{max}(P_t(x)) - \mbox{min}(P_t(x))}{\mbox{max}(P_t(x)) + \mbox{min}(P_t(x))}\mbox{ .}
\end{equation}
Then, the degree of interference at each position and time step is defined as the norm of the numerator of the visibility. For further details on how we calculated this quantity in the quantum walk with fractal disorder see Appendix~\ref{apend:def}.

    \begin{equation}
        \mu(x,t) = |\mbox{max}(P_t(x)) - \mbox{min}(P_t(x))|\mbox{ .}
        \label{eq:degofint}
    \end{equation}

\begin{figure*}[!ht]
    \centering
    \includegraphics[scale = 0.2425]{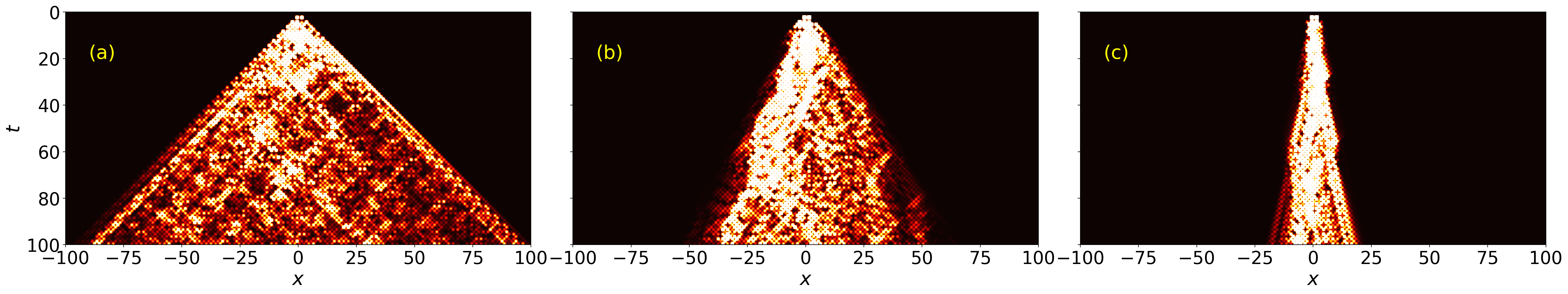}
    \caption{Degree of interference spatial-temporal evolution in the walks following the same parameters of the lower panels (disordered cases) in Fig.~\ref{fig:ca-qw-pxt}, (a) $\theta_H = \theta_F = 15^\circ$, (b) $\theta_H = \theta_F = 45^\circ$ and (c) $\theta_H = \theta_F = 75^\circ$.}
    \label{fig:degofint_equal}
\end{figure*}

In Fig.~\ref{fig:degofint_equal}, we have the degree of interference at each position as a function of time for the same evolution represented at Fig.~\ref{fig:ca-qw-pxt} in the second row, i.e. the disordered case. There, it is possible to see that the interference pattern evolution closely follows that of the probability distribution; in other words, the plot tells us that at each site the interference occurs between its immediate neighbours, as one would expect from the short-range coupling of the quantum walk step evolution.

\begin{figure*}[!ht]
    \centering
    \includegraphics[scale = 0.24]{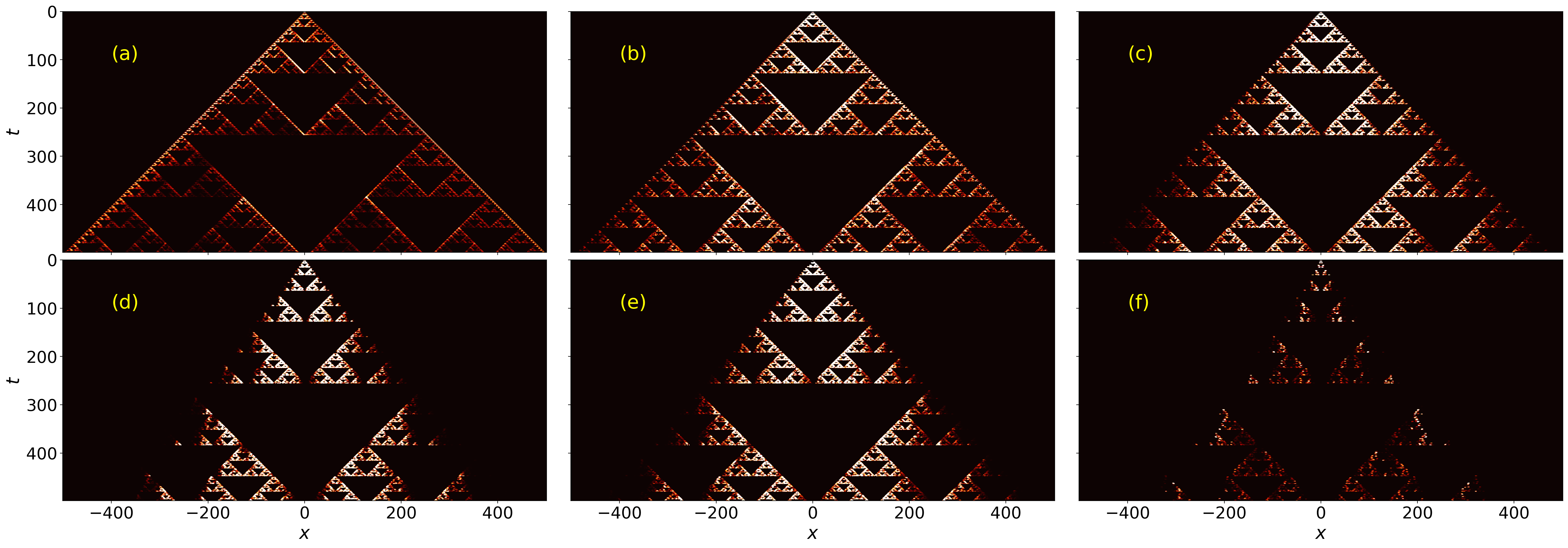}
    \caption{Degree of interference spatial-temporal evolution in walks with $\theta_F = 0^\circ$ initially localized in the origin with coin initial state parameters $\gamma = \pi/2$, $\phi = 0$. In (a) $\theta_H = 5^\circ$, (b) $\theta_H = 15^\circ$, (c) $\theta_H = 30^\circ$, (d) $\theta_H = 45^\circ$, (e) $\theta_H = 60^\circ$ and (f) $\theta_H = 85^\circ$.}
    \label{fig:degofint_thetaH0}
\end{figure*}

By investigating how the degree of interference behaves as a function of the coin operators Eq.~(\ref{Eq:C-explicito}), we also found some interesting patterns that resembles the fractal structure used to implement the disordered coin operator Eq.~(\ref{eq:coinop}) as shown in Fig.~\ref{fig:degofint_thetaH0}. These patterns arise when we set the Fourier coin operator to be equal to the identity, i.e. $\theta_F = 0$, and we start with a coin initial state parameterized by the angles on the Bloch sphere

\begin{equation}
    \psi_0(0) = \cos\frac{\gamma}{2} \ket{U} + e^{i\phi}\sin\frac{\gamma}{2}\ket{D}\mbox{ ,}
    \label{eq:cinstate}
\end{equation}
with angles $\gamma = \pi/2$ and $\phi = 0$. 

One can understand why the pattern resembles a SG fractal by recalling the definition of the degree of interference, Eq.~(\ref{eq:fxt}) and Eq.~(\ref{eq:Adegofint}), with the arithmetic rule used to generate the pattern Eq.~(\ref{eq:fractal}). From Eq.~\ref{eq:fxt} we note that for the degree of interference to be non-zero at a given position at least one of the coin operators applied in its immediate neighbours positions in the previous time instant must be different than the identity operator $C_F(\theta_F = 0)$, i.e. $R^{UD}_{t-1} \ne 0$ and/or $R^{DU}_{t-1} \ne 0$. This matches the sum modulus two where, if $b_{t-1}(x \pm 1)$ are both zero then $b_t(x) = 0$. Moreover, one must have a coin in a superposition state, that is $c_{U}(x + 1, t - 1),c_{D}(x + 1, t - 1) \ne 0$ and/or $c_{U}(x - 1, t - 1),c_{D}(x - 1, t - 1) \ne 0$. The lack of a superposition can come from the application of an identity operator in a state previous state or the application of $C_H$ in one coin at $x + 1$ or $x - 1$. That means when we have $C_H$ acting on both the coin states at $x \pm 1$ the interference degree can be zero or non-zero, based on the prior states at these positions, and when $C_H$ acts on either of them the outcome is the same. That differs from the sum modulus two rule Eq.~\ref{eq:fractal} that states when both $b_{t-1}(x \pm 1)$ are equal to one then $b_t(x) = 0$ and when only one of them is equal to one, $b_t(x) = 1$. Still, we can affirm that the pattern closely resembles the SG fractal for the reason that the majority of points are zero, imposing that the intermediate points in the future are also zero. We also emphasize the dependence of the pattern on the coin initial state (compare Fig.~\ref{fig:degofint_thetaH0}(a) with fig.~\ref{fig:degofint_thetaH0}(f)). Given that one has a small $\theta_H$, then one of the components of the coin state in a given position can be made smaller resulting in a smaller interference for the points that should be zero and a not for the ones that should not. At Appendix~\ref{apend:calc} we show an analytical calculation of the initial steps of the degree of interference evolution at Fig.~\ref{fig:degofint_thetaH0} for any $\theta_H$.

\subsection{Entanglement entropy}

To quantify the amount of entanglement generated through the position-coin system we have employed the entanglement entropy,
\begin{equation}
    S_E \equiv -\mbox{tr}(\rho\log\rho)\mbox{ ,}
\end{equation}
where $\rho$ is the density operator of the system. Here, we focus only on the entropy of the coin subsystem. As we are treating a two-level system, the entanglement entropy varies between $0 \le S_E \le 1$, with zero corresponding to a separable state and with one to a fully entangled state, and it is given by
\begin{equation}
    S_E = - \lambda_+ \log \lambda_+ - \lambda_-\log\lambda_-\mbox{ ,}
\end{equation}
with $\lambda_{\pm }$ being the eigenvalues of the coin density operator $\rho_c$.

Since the evolution of the coin entanglement entropy in a quantum walk often presents an initial increase to then stabilize around a given saturate value, we consider the average entanglement entropy in the asymptotic regime, $t \ge t_0 \gg 1$. The time evolution of the entanglement entropy also depends on the coin initial state and coin operator used. Therefore, we take the initial time after which we can consider the regime as being a quasi-stationary one, $t_0$, based on each evolution.

\begin{figure*}[!ht]
    \centering
    \includegraphics[scale = 0.15]{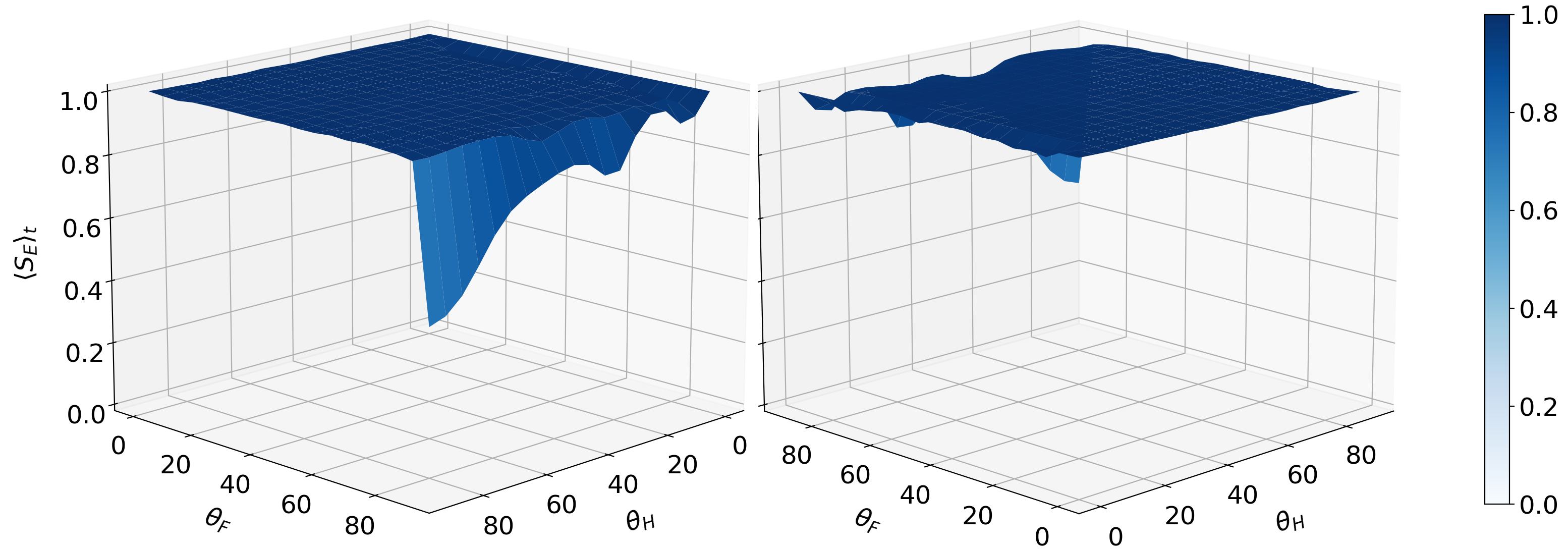}
    \caption{Average coin entanglement entropy in the asymptotic regime as a function of the parameters of the two possible coin operators 
    Eq.~\ref{Eq:C-explicito}, $\theta_H(\mbox{deg})$ and $\theta_F(\mbox{deg})$ respectively. The initial state considered was the one with $\phi = \gamma = \pi/2$ with the initial position in the origin, $x_0 = 0$.}
    \label{fig:se_thetaH_x_thetaF}
\end{figure*}

We start by analyzing the coin entanglement entropy as a function of the coin operator parameters Eq.~(\ref{Eq:C-explicito}), $\theta_H$ and $\theta_F$, at Fig.~\ref{fig:se_thetaH_x_thetaF}. The coin initial state considered is the one with $\phi = \gamma = \pi/2$ in the qubit Bloch-sphere Eq.~(\ref{eq:cinstate}) and localized at the origin. Figure~\ref{fig:se_thetaH_x_thetaF} show us that the entanglement entropy reaches its maximum value for the set of parameters with $\theta_F \lesssim 85^{\circ}$, which is almost the entire set of parameters. For $\theta_F = 90^{\circ}$, the maximum value is obtained with $\theta_H = 0^{\circ}$, after which its start to decay and reach the minimum value of $\langle S_E\rangle_t \approx 0.498$ with $\theta_H = 90^{\circ}$. 
This is an interesting feature since it does not happen in the standard quantum walk evolution. Moreover, a similar effect was reported for other types of disordered quantum walk, with the disorder also in the coin operation and in the shift operator~\cite{vieira2013dynamically,vieira2014entangling,pires2019multiple,buarque2019aperiodic}. However, all of them impose a random quantum walk evolution whereas in the present model the evolution is deterministic. The presence of temporal disorder in the evolution of those walks is the key factor leading to the same phenomenon, as was asserted in ~\cite{vieira2014entangling,naves2022enhancing}.

When we have fixed the coin operators assuming $\theta_H = \theta_F = \pi/4$, we observed that by varying the coin initial state angles the average entanglement entropy does not change for the entire set of parameters. That is indicative of the robustness of the generation of entanglement between the coin and position states with regard to changes in the initial state~\cite{pires2020quantum}.


\subsection{Asymptotic regime}

Next, we probe how the coin state evolves towards its time asymptotic regime. To that, we employ the trace distance measure 

\begin{equation}
    D(\rho,\sigma) = \frac{1}{2}\|\rho - \sigma\|_1\mbox{ ,}
\end{equation}
where $\|A\|_1 = \mbox{tr}(\sqrt{A^\dagger A})$. If a quantum system is evolving in a stationary regime, its density operator does not changes with time, $\rho(t+\Delta t) = \rho(t)$, therefore the trace distance between $\rho(t+\Delta t)$ and $\rho(t)$ would return zero. In this way we see how the trace distance between two time-consecutive states evolves.

\begin{figure*}[!ht]
    \centering
    \includegraphics[scale = 0.24]{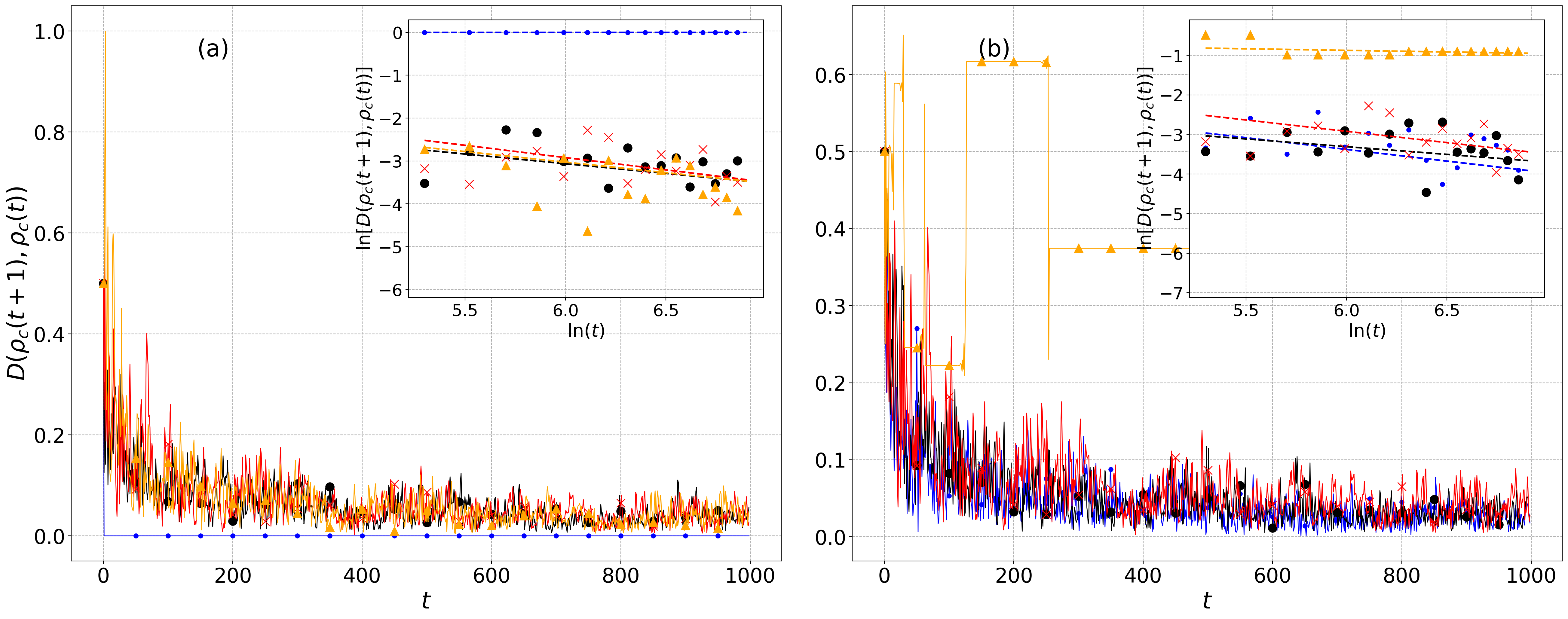}
    \caption{Trace distance between time-consecutive coin states as a function of time for (a) different $\theta_H$, with $\theta_F = \pi/4$, and (b) different $\theta_F$ with $\theta_H = \pi/4$. The different angles used in each plot follows the colors and patterns $0$ (blue dotted), $\pi/6$ (black circled), $\pi/4$ (red crossed) and $\pi/2$ (orange up triangle). In each evolution a localized initial state was considered with coin initial state parameterized by $\phi = \gamma = \pi/2$. The insets indicates the log-log graph of the same plots with linear fittings of the form $-\beta\ln(t) + \alpha$. For (a) we have the following angular and linear coefficients: $\beta = 0.00 \pm 0.00$, $\alpha = 0.00 \pm 0.00$ (blue dotted); $\beta = 0.45 \pm 0.04$, $\alpha = -0.4\pm 0.3$ (black circled); $\beta = 0.57 \pm 0.04$, $\alpha = 0.5 \pm 0.3$ (red crossed); 
    $\beta = 0.49 \pm 0.04$, $\alpha = -0.1 \pm 0.3$ (orange up triangle). For (b) we have the following ones: $\beta = 0.59 \pm 0.05$, $\alpha = 0.2 \pm 0.03$ (blue dotted); $\beta = 0.39 \pm 0.04$, $\alpha = -1.0 \pm 0.3$ (black circled); $\beta = 0.57 \pm 0.04$, $\alpha = 0.5 \pm 0.3$ (red crossed); 
    $\beta = 0.08 \pm 0.04$, $\alpha = -0.39 \pm 0.06$ (orange up triangle). }
    \label{fig:time_td_x_theta}
\end{figure*}

Initially, we fix one of the parameters of the coin operator $\theta_H = \pi/4$ in Fig.~\ref{fig:time_td_x_theta}(a). There, we can see that for all $\theta_F$ but $\theta_F = 0$, the trace distance decays essentially in the same way. The inset gives us the log-log graph of the same curves with the respective linear fittings yielding the decay exponents if $D(\rho_c(t+1), \rho_c(t)) \propto t^{-\beta}$, confirming this observation. 

When we fix the other coin operator with $\theta_F = \pi/4$ Fig.~\ref{fig:time_td_x_theta}(b), the evolution with $\theta_H = 0$ does not decay as fast as in the case where $\theta_F = 0$ in Fig.~\ref{fig:time_td_x_theta}(a), with $\beta \approx 0.59$. For the remaining angles, the decay rates are essentially the same with the exception of $\theta_H = \pi/2$ that gives us $\beta \approx 0.08$. In this angle something interesting happens, the trace distance does not decay, on average, monotonically, having intervals of constant evolution with some increases in between.

\begin{figure*}[!ht]
    \centering
    \includegraphics[scale = 0.24]{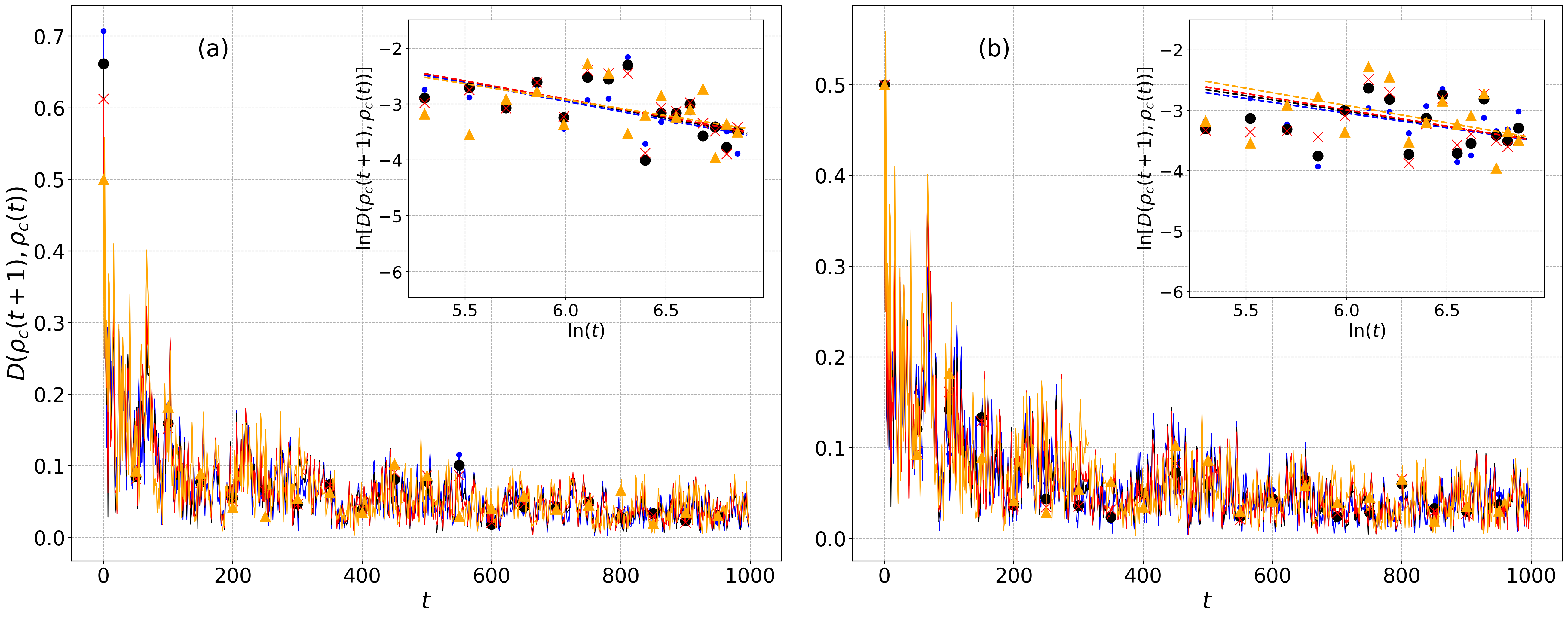}
    \caption{Trace distance between time-consecutive coin states as a function of time for different angles of the coin initial state Bloch sphere $\phi$ (a), with $\gamma = \pi/2$, and different $\gamma$(b) with $\phi = \pi/2$ (see eq.~\ref{eq:cinstate}). The different angles used in each plot follows the colors and patterns $0$(blue dotted), $\pi/6$(black circled), $\pi/4$(red crossed) and $\pi/2$(orange up triangle). In each evolution a localized initial state was considered with coin operators given by $\theta_H = \theta_F = \pi/4$. The insets indicates the log-log graph of the same plots with linear fittings of the form $-\beta\ln(t) + \alpha$. For (a) we have the following angular and linear coefficients: $\beta = 0.67 \pm 0.04$, $\alpha = 1.1 \pm 0.3$ (blue dotted); $\beta = 0.66 \pm 0.04$, $\alpha = 1.0 \pm 0.3$ (black circled); $\beta = 0.65 \pm 0.04$, $\alpha = 1.0 \pm 0.2$ (red crossed); 
    $\beta = 0.57 \pm 0.04$, $\alpha = 0.5 \pm 0.3$ (orange up triangle). For (b) we have the following ones: $\beta = 0.48 \pm 0.05$, $\alpha = -0.1 \pm 0.3$ (blue dotted); $\beta = 0.51 \pm 0.04$, $\alpha = 0.0 \pm 0.3$ (black circled); $\beta = 0.54 \pm 0.04$, $\alpha = 0.2 \pm 0.3$ (red crossed); 
    $\beta = 0.57 \pm 0.04$, $\alpha = 0.5 \pm 0.3$ (orange up triangle).}
    \label{fig:time_td_x_cinstate}
\end{figure*}

Fig.~\ref{fig:time_td_x_cinstate} shows the time evolution of the trace distance between two consecutive coin states when we set the coin operators to be the those with $\theta_H = \theta_F = \pi/4$ and change the coin initial state. In Fig.~\ref{fig:time_td_x_cinstate}(a), the polar angle is set $\gamma 
= \pi/2$ while we change the phase angle, $\phi$. We see that changes in this parameter does not changes the trace distance decay behavior, on average, giving us essentially the same decay exponent $\beta \approx 0.66$ for $\phi = \{0,\pi/6,\pi/4\}$ with the most divergent, but still very close, $\beta \approx 0.57$ for $\phi = \pi/2$. Setting $\phi = \pi/2$ and changing $\gamma$, Fig.~\ref{fig:time_td_x_cinstate}(b) demonstrates that changes in this parameter also does not change the trace distance decay significantly, now even when we set $\gamma = \pi/2$.

These results tell us that despite our quantum walk model being disordered, it goes to the quasi-stationary regime essentially in the same way as if we had fixed the coin operators used and changed the coin initial state and vice-versa, excluding the cases where the parameters are near to $\pi/2$. This is strongly indicative that this property derives from the quantum walk protocol and only has a small limited dependence on the initial conditions. In the standard Hadamard walk, the pace at which the coin state converges to the stationary regime depends on the initial state and coin operator used. For the initially localized and balanced coin state the decay exponent is $\beta = 3/2$, as opposed to the average decay exponent that we obtained $\beta \approx 0.55$. Comparing with other types of disordered quantum walks, we see that the introduction of a fractal disorder in the coin operator induces a faster transition to the stationary regime than when one introduces a dynamically and fluctuating randomness~\cite{vieira2013dynamically,vieira2014entangling}, where the decay exponent $\beta = 1/4$. The decay exponent of the walk with fractal disorder is also greater than the ones in the generalized elephant quantum walk, where the randomness is on the step sizes and selected according to the $q$-exponential, when its parameter $q$ is less than one but greater than half, $\beta(q) \approx 1/4$, but in the uniform distribution limit $q \rightarrow \infty$ it becomes smaller, where $\beta(q) \approx 0.76$.

\section{Discussion}
\label{sec:discussion}

Within the context of Quantum Walks, the disorder of our SG case befalls twofold: every site of the grid is associated to a given coin protocol that is updated at each time step according to the dynamics Eq.~(\ref{eq:fractal}). On the one hand, it is know that the slightest spatial disorder induces an absence of diffusion and the subsequent localization of the quantum walker wave; on the other hand, temporal randomness can actually enhance spreading of the wavepacket, especially when such changes occur at typically long time scales~\cite{schreiber2011decoherence}. Thus, taking into consideration this model has got spatial disorder, we would expect a dramatic decrease of the exponent $\alpha $, which is not the case. As shown, the system preserves a level of diffusion not far from the standard ballistic behavior with mild dependence on the angle of the coin. In order to understand that, we have inspected the local statistics of coin operator updates. Excluding the central grid site, $x=0$, we have verified that the average time for coin change, $\tau _x$, is never less than a two-digit figure, with that average time increasing with $|x|$, namely $\tau _5 = 74$, $\tau _{30} = 146$,  $\tau _{100} = 219$. Combining both elements, we understand the long characteristic scale of the coin dynamics manages to zero out to a large extent the likely localization that would be induced by static disorder. At the same time, we have verified the effect of spatial disorder in augmenting quantum entanglement~\cite{buarque2019aperiodic} (excluding the particular coin angles such as $\theta = 90$), which is lessened by the temporal disorder effects of the dynamics Eq.~(\ref{eq:fractal}). In other words, we have verified that deterministic fractal disorder enhances quantum entanglement without achieving the maximal value though. It is precisely that deterministic nature of the self-similar  fractal dynamics that reins quantum entanglement strengthening in, as previously hinted~\cite{vieira2013dynamically}.   
That said, we can particularly envisage this type of architecture as a strong candidate for tailored problems for which one aims at increasing entanglement whilst ensuring high level of spreading, e.g., search algorithms in a quantum cryptography environment~\cite{abd2019encryption}. 

\section{Final remarks}
\label{sec:remarks}

We have studied quantum walks with spatiotemporal fractal disorder, employing the paradigmatic Sierpinski gasket as a prototype. Throughout our investigation, we have explored multiple facets of the quantum walk, providing insights into its transport properties, degree of interference, entanglement entropy, and trace distance between time-consecutive coin states.

The employment of fractal principles in the creation of patterned structures is an avenue of research to be studied from both fundamental and applied perspectives. In this manuscript, we have provided a novel application of fractal geometry, namely, we showed that the Sierpinski gasket can be employed to build photonic architectures aimed at realizing highly nontrivial site- and time-dependent quantum dynamics.  While fractal structures were previously addressed in the scope of QWs
~\cite{agliari2010quantum,berry2010quantum,patel2012search,tamegai2018spatial,sato2020scaling,xu2020shining,lara2013quantum,andrade2018discrete,agliari2008dynamics,darazs2014transport} those efforts had  different frameworks and goals than what we presented here. Accordingly, with this work where the disorder is deterministically established, we have fulfilled a gap on the characterization of the impact of different sorts of disorder on the properties of a quantum walk.

An important advantage of this work in relation to previous proposals~\cite{berry2010quantum,patel2012search,lara2013quantum,tamegai2018spatial,sato2020scaling} lies in the fact that the versions of the discrete-time QWs on fractals studied in those works require the use of high-dimensional coin operators that are harder to implement. Our protocol simply considers a two-state coin. 
Specifically, in Ref.~\cite{berry2010quantum} it was used a $d$-dimensional coin Hilbert space. In Refs.~\cite{patel2012search,lara2013quantum,tamegai2018spatial,sato2020scaling} their flip-flop walk requires a four-dimensional Grover coin. Thus, the experimental feasibility of the present QW dynamics  presented a clear implementable advantage; furthermore, it can be realized with the state-of-art photonics technology. Still, with minor changes in the apparatus described in Ref.~\cite{wang2018dynamic}, it is possible to implement the fractal concatenation of optical elements where the role of time $t$ is played by the propagation direction. The nonrandom character of our  proposal has the additional feasibility of not requiring an extra sampling processing over random disorder.

The present work offers new alternatives for bringing the fields of fractal and photonics even more closer~\cite{segev2012fractal,graydon2019fractal}.
Still, this contribution brings fresh insights into the physics of disordered photonics~\cite{wiersma2013disordered}. An interesting open problem to study is how the wavepacket properties behave when a multifractal structure  is employed to incorporate disorder in the system.

\begin{acknowledgments}
CBN acknowledges the prospective funding from the Department of Physics of Stockholm University. SMDQ thanks CNPq (Grant 302348/2022-0) for financial support. DOSP acknowledges the support by the Brazilian funding agencies CNPq (Grant No. 304891/2022-3), FAPESP (Grant No. 2017/03727-0), and the Brazilian National Institute of Science and Technology of Quantum Information (INCT/IQ).  
\end{acknowledgments}

 \appendix
\section{The degree of interference}
\subsection{Definition}
\label{apend:def}
    In the following, we present the way we have calculated the degree of interference in connection with the proposed quantum walk model. As we mentioned previously, the degree of interference is defined as the norm of the quantity responsible for the interference in the visibility. The visibility, suitable for probability waves, is 

    \begin{equation}
        \mathcal{V} = \frac{P_{max} - P_{min}}{P_{max} + P_{min}}\mbox{ .}
    \end{equation}
    Then, the degree of interference at each position and time step is defined as the norm of the numerator of the visibility

    \begin{equation}
        \mu(x,t) = |P_{max}(x,t) - P_{min}(x,t)|\mbox{ .}
    \end{equation}

    Let $\ket{\psi(t)}$ be the walker state at time $t$ expanded as

    \begin{equation}
        \Psi(t) = \sum_x \ket{x} \otimes (
        c_{U}(x,t)\ket{U} + c_{D}(x,t)\ket{D})\mbox{ .}
    \end{equation}

    The coin operator $\hat{R}_t$ in the quantum walk is defined following the Sierpinski Gasket fractal rule so that

    \begin{equation}
        \hat{R}_t = \sum_x \ketbra{x}{x}\otimes\hat{R}_t(x)\mbox{ ,}
    \end{equation}
    with $\hat{R}_t(x)$ given by eq.~(\ref{eq:coinop}).

    Therefore, the recursive maps that provides the walker state coefficients' time evolution now are going to include the space-time dependent coin operator factors. Let 

    \begin{equation}
        \hat{R}_t(x) = R_t^{UU}(x)\ketbra{U} + 
        R_t^{UD}(x)\ketbra{U}{D} +
        R_t^{DU}(x)\ketbra{D}{U} +
        R_t^{DD}(x)\ketbra{D}\mbox{ ,}
    \end{equation}
    consequently, the recursive map for the state coefficients is going to be, assigning the up state to a displacement to the right

    \begin{align}
        c_{U}(x,t+1) &= R_t^{UU}(x-1)c_{U}(x-1,t) +
        R_t^{UD}(x-1)c_{D}(x-1,t)\label{eq:cup} \\
        \nonumber \\
        c_{D}(x,t+1) &= R_t^{DU}(x+1)c_{U}(x+1,t) +
        R_t^{DD}(x+1)c_{D}(x+1,t)\mbox{ .}
        \label{eq:cdown}
    \end{align}

    The position probability distribution is given by the square norm of the walker state's coefficients

    \begin{equation}
        P(x,t) = |c_{U}(x,t)|^2 + |c_{D}(x,t)|^2\mbox{ .}
    \end{equation}

    Using the recursive relations Eqs.(\ref{eq:cup})~(and)~(\ref{eq:cdown}), we find

    \begin{align}
        P(x,t) &= |R_{t-1}^{UU}(x-1)c_{U}(x-1,t-1) +
        R_{t-1}^{UD}(x-1)c_{D}(x-1,t-1)|^2 + \nonumber \\
        &+ |R_{t-1}^{DU}(x+1)c_{U}(x+1,t-1) +
        R_{t-1}^{DD}(x+1)c_{D}(x+1,t-1)|^2 \\
        &= |R_{t-1}^{UU}(x-1)c_{U}(x-1,t-1)|^2 + 
        |R_{t-1}^{DD}(x+1)c_{D}(x+1,t-1)|^2 
        |R_{t-1}^{UD}(x-1)c_{D}(x-1,t-1)|^2 + \nonumber \\
        &+ |R_{t-1}^{DU}(x+1)c_{U}(x+1,t-1)|^2 + |R_{t-1}^{UD}(x-1)c_{D}(x-1,t-1)|^2 +       \nonumber \\
        &+ R_{t-1}^{UU}(x-1)(R_{t-1}^{UD}(x-1))^*
        c_{U}(x-1,t-1)c_{D}(x-1,t-1)^* + \nonumber \\
        &+ R_{t-1}^{DU}(x+1)(R_{t-1}^{DD}(x+1))^*
        c_{U}(x+1,t-1)c_{D}(x+1,t-1)^* + \mbox{c.c.}
    \end{align}

    Defining the complex part as a function $f(x,t)$:
    
    \begin{align*}
        f(x,t) = 
        R_{t-1}^{UU}(x-1)(R_{t-1}^{UD}(x-1))^*
        c_{U}(x-1,t-1)c_{D}(x-1,t-1)^* + \nonumber \\
        + R_{t-1}^{DU}(x+1)(R_{t-1}^{DD}(x+1))^*
        c_{U}(x+1,t-1)c_{D}(x+1,t-1)^*
        \label{eq:fxt}
    \end{align*}

    the maximum probability and the minimum probability are going to be

    \begin{align}
        P_{\max}(x,t) &= |R_{t-1}^{UU}(x-1)c_{U}(x-1,t-1)|^2 + 
        |R_{t-1}^{UD}(x-1)c_{D}(x-1,t-1)|^2 
        +  \nonumber \\ 
        &+ |R_{t-1}^{DU}(x+1)c_{U}(x+1,t-1)|^2 + |R_{t-1}^{DD}(x+1)c_{D}(x+1,t-1)|^2 + \nonumber \\
        &+ f(x,t) + f^*(x,t)\mbox{ ,}
    \end{align}
    
    \begin{align}
        P_{\min}(x,t) &= |R_{t-1}^{UU}(x-1)c_{U}(x-1,t-1)|^2 + 
        |R_{t-1}^{UD}(x-1)c_{D}(x-1,t-1)|^2 + \nonumber \\
        &+ |R_{t-1}^{DU}(x+1)c_{U}(x+1,t-1)|^2 + |R_{t-1}^{DD}(x+1)c_{D}(x+1,t-1)|^2 -
        f(x,t) - f^*(x,t)\mbox{ .}
    \end{align}

    Therefore, the degree of interference will be given by

    \begin{equation}
        \mu_{x,t} = |4\mbox{Re}(f(x,t))|\mbox{.}
        \label{eq:Adegofint}
    \end{equation}

\subsection{Degree of interference evolution}
\label{apend:calc}

In this section, we move on to calculate the initial time steps of the evolution used to plot the degree of interference patterns in fig.~\ref{fig:degofint_thetaH0}. In these evolution the coin operators used are $\hat{C}_H(\theta_H)$ and the identity operator, $\hat{C}_F(\theta_F = 0) = \mathbb{I}_{2 \times 2}$.

The initial state used is:

\begin{equation}
    \Psi_0 = \ket{0}\otimes\frac{\ket{\uparrow} + \ket{\downarrow}}{\sqrt{2}}\mbox{ .}
\end{equation}

At time $t = 0$ the fractal pattern tells us that the coin operator is equal to $\hat{C}_H$ at $x = 0$ and the identity for $x \ne 0$. Therefore, after the application of the unitary operator we are going to have

\begin{equation}
    \Psi_1 = \frac{(\cos\theta_H + \sin\theta_H)}{\sqrt{2}}\ket{+1,\uparrow} + \frac{(\sin\theta_H - \cos\theta_H)}{\sqrt{2}}\ket{-1,\downarrow}\mbox{ .} 
\end{equation}

For $ t = 1$, the coin operator follows $\hat{R}_1 = \{\hat{C}_H, x = \pm 1; \mathbb{I}, x \ne \pm 1\}$. Then, by applying it to the state following with the application of the shift operator:

\begin{align*}
    \Psi_2 &= \frac{\cos\theta_H}{\sqrt{2}}(\cos\theta_H + \sin\theta_H)\ket{+2,\uparrow} 
    + \frac{\sin\theta_H}{\sqrt{2}}\ket{0}\otimes[(\sin\theta_H - \cos\theta_H)\ket{\uparrow} + (\cos\theta_H + \sin\theta_H)\ket{\downarrow}] - \\ 
    &- \frac{\cos\theta_H}{\sqrt{2}}(\sin\theta_H - \cos\theta_H)\ket{-2,\downarrow}\mbox{ .}
\end{align*}

When $t = 2$ and $t = 3$, $\hat{R}_2 = \{\hat{C}_H, x = \pm 2; \mathbb{I}, x \ne \pm 2\}$ and $\hat{R}_3 = \{\hat{C}_H, x = \pm 1, \pm 3; \mathbb{I}, x \ne \pm 1, \pm 3\}$. The states at these time are, respectively:

\begin{align*}
    \Psi_3 &= \frac{\cos^2\theta_H}{\sqrt{2}}(\cos\theta_H + \sin\theta_H)\ket{+3,\uparrow} 
     + \frac{\cos^2\theta_H}{\sqrt{2}}(\sin\theta_H - \cos\theta_H)\ket{-3,\downarrow} +   \nonumber \\
    &+ \frac{\sin\theta_H}{\sqrt{2}}\ket{+1}\otimes[(\sin\theta_H - \cos\theta_H)\ket{\uparrow} + \cos\theta_H(\cos\theta_H + \sin\theta_H)\ket{\downarrow}] -  \nonumber  \\
    &- \frac{\sin\theta_H}{\sqrt{2}}\ket{-1}\otimes[\cos\theta(\sin\theta_H - \cos\theta_H)\ket{\uparrow} - (\cos\theta_H + \sin\theta_H)\ket{\downarrow}] \mbox{ , }
\end{align*}

\begin{align*}
    \Psi_4 &= \frac{\cos^3\theta_H}{\sqrt{2}}(\cos\theta_H + \sin\theta_H)\ket{+4,\uparrow} - \frac{\cos^3\theta_H}{\sqrt{2}}(\sin\theta_H - \cos\theta_H)\ket{-4,\downarrow} + \\
    &+ \frac{\sin\theta_H}{\sqrt{2}}\cos\theta_H\ket{+2}\otimes\{[\sin\theta_H(1+\sin\theta_H) + \cos\theta_H(\sin\theta_H - 1)]\ket{\uparrow} + \cos\theta_H(\cos\theta_H + \sin\theta_H)\ket{\downarrow}\} + \\
    &+ \frac{\sin\theta_H}{\sqrt{2}}\ket{0}\otimes\{[\sin\theta_H(\cos\theta_H + \sin\theta_H) - \cos^2\theta_H(\sin\theta_H - \cos\theta_H)]\ket{\uparrow} + \\ 
    & \qquad\qquad\qquad  + [\sin\theta_H(\sin\theta_H - \cos\theta_H) - \cos^2\theta_H(\cos\theta_H + \sin\theta_H)]\ket{\downarrow}\} + \\
    &+ \frac{\sin\theta_H}{\sqrt{2}}\cos\theta_H\ket{-2}\otimes\{\cos\theta_H(\sin\theta_H - \cos\theta_H)\ket{\uparrow} - [\sin\theta_H(\sin\theta_H + 1) + \cos\theta_H(\sin\theta_H - 1)]\ket{\downarrow}\} \mbox{.}
\end{align*}

Now we are going to use these states and the coin operators defined for each time step to calculate the degree of interference. At $t = 0$, as we do not have a previous time step, we set the interference degree to be equal to one at the origin:

\begin{equation}
    \mu(x,0) =  \begin{cases}
                    1,\; x = 0 \\
                    0,\; x \ne 0
                \end{cases}
\end{equation}

At $t = 1$, we recall that for the degree of interference to be non-zero at a given position the coin operator applied in its immediate neighbours must be $\hat{C}_H$ and the coin has to be in a superposition as well. For $x = \pm 1$ these requirements are satisfied since the initial state located at the origin is in a superposition and $\hat{C}_H$ was applied. Then, we obtain:

\begin{equation}
    \mu(x,1) =  \begin{cases}
                    2\sin\theta_H\cos\theta_H,\; x = \pm 1 \\
                    0,\; x \ne \pm 1
                \end{cases}
\end{equation}

In the next time step, $t = 2$, given that the state is an entangled one, the requirement of having a superposition of coin states in a position is not fulfilled. Therefore, the interference degree is zero for every position:

\begin{equation}
    \mu(x,2) =  0,\; \forall x\mbox{ .}
\end{equation}
Here we can already notice a difference between the degree of interference pattern and the Sierpinski-Gasket fractal. 

For $t = 3$ the same happens, however for the reason that at $x = 0$ the coin operator used is the identity, while at $x = \pm 2$ $\hat{C}_H$ is used but the coin states are not in superposition:

\begin{equation}
    \mu(x,3) =  0,\; \forall x\mbox{ .}
\end{equation}

At $t = 4$ we have to look at the walker state and the coin operator at $t = 3$. We notice that at $x = \pm 1$ the coin state is in a superposition and the identity is not applied at these positions. Therefore, the only points that the interference degree can be non-zero are at $x = 0$ and $x = \pm 2$. Performing the calculations, we find:

\begin{equation}
    \mu(x,4) =  \begin{cases}
                    4\cos^2\theta_H\sin^3\theta_H\cos2\theta_H,\; x = 0 \\
                    2\cos^2\theta_H\sin^3\theta_H\cos2\theta_H,\; x = \pm 2 \\
                    0,\; x \ne 0, \pm 2
                \end{cases}
\end{equation}

Taking one more final step, $t = 5$, we find once again that the interference degree is zero for every position given that identity operator is applied to every position but at $x = \pm 4$, however the coin state at these positions are not in superposition:

\begin{equation}
    \mu(x,5) =  0,\; \forall x\mbox{ .}
\end{equation}

%
\end{document}